\documentclass[structabstract]{aa}
\usepackage{txfonts}
\usepackage{graphicx}
\usepackage{natbib}
\bibpunct{(}{)}{;}{a}{}{,} 

\newcommand{\dx}{\mathrm{d}}
\def\phx{{\texttt{PHOENIX}}}

\begin{document}

\title{Theoretical light curves of type Ia supernovae} 
\author{D. Jack\inst{1}
  \and P. H. Hauschildt\inst{1} 
   \and E. Baron\inst{1,2}} 

\institute{Hamburger Sternwarte, Gojenbergsweg 112, 21029 Hamburg, Germany\\
  e-mail: djack@hs.uni-hamburg.de; yeti@hs.uni-hamburg.de
  \and Homer L. Dodge Department of Physics and Astronomy, University of Oklahoma, 440 W Brooks, Rm 100, Norman, OK 73019-2061 USA\\
  e-mail: baron@ou.edu}

\date{Received 13 April 2010 /
      Accepted 9 February 2011}

\abstract {}
{We present the first theoretical SN Ia light curves calculated with
the time-dependent version of the general purpose model atmosphere code \phx.
Our goal is to produce light curves and spectra of
hydro models of all types of supernovae.}
{We extend our model atmosphere code {\phx} to calculate type Ia supernovae light curves.
A simple solver was implemented which keeps track of
energy conservation in the atmosphere during the free expansion phase.}
{The correct operation of the new additions to {\phx} were verified in test
calculations. Furthermore, we calculated theoretical light curves and compared
them to the observed SN Ia light curves of SN 1999ee and SN 2002bo. We obtained LTE as well as
NLTE model light curves.}
{We have verified the correct operation of our extension into the time domain.
We have calculated the first SN Ia model light curves using \phx\ in both LTE and NLTE.
For future work the infrared model light curves need to be further investigated.}


\maketitle

\section{Introduction}

All types of supernovae are important for the role that they play in
understanding stellar evolution, galactic nucleosynthesis, and as
cosmological probes.
Type Ia supernovae are of particular cosmological interest, e.g., because the
dark energy was discovered with Type Ia supernovae
\citep{riess_scoop98,perletal99}.

In dark energy studies, the goal now is to characterize the nature of
the dark energy as a function of redshift. While there are other
probes that will be used (gravitational lensing, baryon acoustic
oscillations), a JDEM or Euclid mission will likely consider supernovae
in some form. In planning for future dark energy studies both from space
and from the ground, it is important to know whether the mission will
require spectroscopy of modest resolution, or whether pure imaging or
grism spectroscopy will be adequate. Several purely spectral
indicators of peak luminosity have been
proposed
\citep{nugseq95,hach06,bongard06a,bronder08,ffj08,ledu09}.
What is required is an empirical and theoretical comparison of both light curve
shape luminosity indicators
\citep{pskovskii77,philm15,rpk96,goldhetal01} and spectral indicators.

To make this comparison one needs to know more about the
physics going on in a supernova explosion and to be able to calculate
light curves and spectra self-consistently.
Thus, we need to extend our code to time-dependent problems.
While our primary focus is on Type Ia supernovae, the time-dependent
radiative transfer code is
applicable to all types of
supernovae, as well as to other objects, e.g., stellar pulsations.

In the following we present the methods we used to implement time
dependence. First we focus on solving the energy equation (first law)
and then present our first theoretical LTE light curves of type Ia supernova
events. In a further section, we present results of NLTE model light
curve calculations.

\section{Energy solver}

In our previous paper \citep{jack09},
we presented our approach to calculate light curves of type Ia
supernovae events. We implemented a simple energy solver into our
general purpose model atmosphere code \phx\ 
where we kept track of the overall energy change of the radiating fluid and the
energy exchange between the matter and radiation.
Simple test light curves confirmed that our approach worked correctly.
We now present a new approach to calculate theoretical light curves of SNe Ia.
Again, we are using a simple solver where we keep track of the energy conservation.
In the new approach, we consider only the energy density of the material.
We assume free expansion and do not solve the equations of hydrodynamics.
In order to obtain the result for the new time step, we use an explicit scheme.
We apply this scheme and compute time steps until radiative equilibrium is reached.
These are the points in our light curves.

The direct change of the energy density of the material considering
absorption and emission of radiation
and energy deposition by gamma rays  is given by
equation (96.7) in \citet[]{found84}
\begin{equation}
\rho\left[\frac{\dx e}{\dx t}+p\frac{\dx}{\dx t}\left(\frac{1}{\rho}\right)\right]=
\int\left(c\chi E-4\pi\eta\right)+\rho\epsilon,
\end{equation}
where $\rho$ is the density, $p$ the gas pressure and
$e$ is the energy density of the material.
The quantities of the radiation field are $\chi$ which is the absorption coefficient,
$\eta$ is the emission coefficient and 
$E=\frac{4\pi}{c}J$ is the radiation energy density with the mean intensity $J$.
To obtain $E$, the radiative transfer equation in spherical symmetry is solved including special relativity.
All additional energy sources are put in $\epsilon$ such as the energy input from gamma ray deposition.
This equation represents the first law of thermodynamics for the material.
The change of the energy density of the material
depends on the coupling of matter and radiation field, the absorption of gamma-radiation and positron annihilation
energy, the expansion work and the change of the ionization and excitation of matter.

Dividing equation 1 by the material density $\rho$ we obtain
\begin{equation}
\frac{\dx e}{\dx t}=
\frac{1}{\rho}\int\left(c\chi E-4\pi\eta\right)\dx \lambda-p \frac{\dx}{\dx t}\frac{1}{\rho}+\epsilon.
\end{equation}
The radiation energy density is given by $E=\frac{4\pi}{c}J$, where $J$ is the mean intensity.
Using this, we obtain for the radiation term
\begin{equation}
\int(c\chi E-4\pi\eta)\dx\lambda=4\pi\int(\chi J-\eta)\dx\lambda=4\pi\int(\chi(J-S))\dx\lambda,
\end{equation}
where $S=\frac{\eta}{\chi}$ is the source function. All these quantities can be derived from the
solution of the radiative transfer equation.
The term of the change of the energy density by the radiation is therefore given by
\begin{equation}
Q=\int\chi_{\lambda}(J_{\lambda}-S_{\lambda})\dx\lambda.
\end{equation}

Another change of the energy density is due to the work $W$ done by the adiabatic expansion
of the SN Ia atmosphere. For a discrete step this work is given by
\begin{equation}
W=p\left(\frac{1}{\rho_{2}}-\frac{1}{\rho_{1}}\right).
\end{equation}
The expansion is assumed to be  homologous.
Since we solve the energy equation for the matter, we do not
include the radiation pressure work. Since the system is radiation
dominated there is the possibility of numerical inaccuracies in
coupling the matter and radiation only by $Q$.
For the calculation of the new radii and densities
as well as a discussion about the accuracy of this assumption see \citet{jack09}.

Including all energy changing effects, the new energy density of the material $e_{2}$ after a discrete time step $\Delta t$ is
explicitly given by
\begin{equation}
e_{2} = e_{1} - p\left(\frac{1}{\rho_{2}}-\frac{1}{\rho_{1}}\right) + \frac{4\pi}{\rho}\Delta t \int
\chi (J-S) \dx \lambda + \epsilon \Delta t,
\label{eq:newenergy}
\end{equation}
where $e_{1}$ is the old energy density.

To obtain all the needed quantities we have to solve the spherically symmetric special relativistic
radiative transfer equation.
The advantage of our new approach is that we do not have to iterate for each time step.
In our previous paper
the calculation of each time step involved an iteration process to obtain the new
matter temperature. With the new approach we can calculate the 
new temperature for the next time step directly from all the known quantities.

The translational energy density of the material is given by
\begin{equation}
e_{trans}=\frac{3}{2}\frac{p}{\rho}=\frac{3}{2}\frac{R}{\mu}T=\frac{3}{2}\frac{N_{A}kT}{m},
\end{equation}
with the mean molecular weigh $\mu$ and the universal
gas constant $R$. The gas pressure is represented by $p$ and the
density by $\rho$. $T$ stands for the temperature of the material.
This equation of the energy density is now used to determine the new temperature after
the next time step.

During the first phase of the SN Ia envelope evolution, the material of the atmosphere is hot and,
therefore, highly ionized. The energy change due to
ionization and excitation changes of the atoms present in the SN Ia atmosphere cannot be neglected.
\phx\ already solves the equation of state (EOS), where all the excitation and ionization stages
of the present atoms and molecules are included. Using the EOS, we obtain the overall energy density of the material
by the sum of the ionization energy $e_{ions}$ and the translational energy $e_{trans}$
\begin{equation}
e=e_{trans}+e_{ions}.
\end{equation}

Hence, the energy density change of the material goes into a change of the translational energy
and the ionization energy, which both depend on the temperature.
Therefore, we obtain the new temperature by an iteration scheme.
The matter density at the next point in time is determined by homologous expansion.
A first temperature guess is used, and the EOS is solved to obtain the ionization energy density.
Combined with the translational energy density, the overall energy density is computed.
This is checked against the target energy density, which we obtained equation \ref{eq:newenergy}.
If the obtained energy density is incorrect,
a new temperature guess is made. This new temperature guess is obtained by assuming a linear
dependence of the energy density and temperature. The current temperature guess is iterated to the target energy density.
It takes about 5 - 10 iteration steps to determine the new temperature.
If the EOS delivers the correct target energy density,
the new temperature of the next time step has been found.
The accuracy relative of the energy density in this iteration process is set to $10^{-5}$.

\subsection{$\gamma$-ray deposition}

The maximum of the light curve of an SN Ia event is observed around 20 days after  explosion.
Causing this later maximum of the light curve of an SN Ia event is the energy release into the envelope caused by
the radioactive decay of $^{56}$Ni and its also radioactive decay product $^{56}$Co.
Therefore, this energy deposition has a strong influence on the energy change of the SN Ia envelope structure.
Hence, the energy deposition due to radioactive decay has to be taken into account
for the calculation of the SN Ia envelope evolution.

The energy deposition due to the $\gamma$-rays emitted by radioactive isotopes needs to be computed by a radiative
transfer solver for the $\gamma$-rays.
In this work, we solve the $\gamma$-ray deposition with the assumption of a gray atmosphere for the $\gamma$-rays.
\citet{jeffery98} did a detailed study of the $\gamma$-ray deposition and pointed out
that this is an adequate approach to calculate $\gamma$-ray deposition in SN Ia atmospheres.
In the decay of a $^{56}$Ni nucleus, a $\gamma$-photon is emitted with an energy of 2.136~MeV. The $^{56}$Co nucleus
decays to
an $^{56}$Fe nucleus and emits a $\gamma$-photon, which has an energy of 4.566~MeV.
In the decay of $^{56}$Co about 19\% of the energy is released by positrons. 
The positrons are assumed to be locally trapped. They annihilate by emitting two photons each with an energy of 512~keV,
which has to be taken into account for the energy deposition calculation.
The opacity is considered to be constant and a pure absorption opacity, meaning that no scattering is assumed.
As in \citet{jeffery98}, $\kappa_{\gamma}=0.06<Z/A> {\rm cm}^{2} {\rm
  g}^{-1}$ was chosen as the opacity. $<Z/A>$ is the proton
fraction, which is counts both bound and free electrons, since the
electron binding energy is small compared to the energy of gamma rays.
The energy deposition into the atmosphere per unit time is given by
\begin{equation}
\epsilon=4\pi\frac{\chi}{\rho} J,
\end{equation}
where $J$ is the mean intensity, which has been obtained by solving the gray radiative transfer for the
$\gamma$-rays.
This obtained energy deposition has to be taken into account for the calculation of the overall energy change.

\subsection{Adaptive time step procedure}

The timescale for energy changes in the SN Ia envelope
will change during the evolution of the light curve.
In order to save computation time, the light curves have to be calculated
with the optimal time step size for each phase of light curve evolution.
Therefore, we implemented an adaptive time step routine to determine the optimal time step
size for the current time step.

The energy change $\Delta e$ of the energy of the material $e$ may be approximated by
\begin{equation}
\Delta e=x\cdot e=\Delta t \cdot \left( Q + \epsilon\right),
\label{eq:adapt}
\end{equation}
where Q is the energy change of the interaction with the radiation,
and $\epsilon$ is the energy deposition by the gamma rays. The energy change due to
the expansion is ignored in this case. On the one hand, this energy change depends on the new matter density after the
time step, which is unknown because it depends on the size of the time step itself.
Furthermore, the energy change because of the expansion is small compared to the changes caused by the energy transport and
the energy deposition by $\gamma$-rays.
The idea of the adaptive time step procedure is to limit the energy change to a prescribed amount of the energy of the material.
Thus, rewriting equation \ref{eq:adapt}, we obtain the time step size $\Delta t$ for the current time step by
\begin{equation}
\Delta t=\frac{e}{Q+\epsilon}\cdot x,
\end{equation}
where $x$ is the introduced limiting energy change factor. The factor $x$ ranges between $x_{min}$
and $x_{max}$, which mark the largest and smallest allowed energy change. These are input parameters
for the adaptive time step procedure. The time step size is calculated for every layer,
and the minimum time step size of all layers is used for the energy solver.

Each time the adaptive time step procedure is called, it checks if the energy changes of the time step before
were too big or could have been bigger.
If the minimum time step size is in a different layer,
the same value of $x$ is kept for the next time step.
If the minimum is in the same layer and the sign of the energy change does not change,
the previous time step might have been too small. Thus, the factor $x$ is increased for the following time step.
If the sign changes, the last time step might have been too large, therefore, the factor $x$ is decreased.
This means that for each time step, the allowed energy change is adapted and the factor $x$ is updated to get the optimal time step during the whole evolution of the SN Ia atmosphere.

\section{Test calculations}

All new implemented physical processes of the simple energy solver have to be tested.
For the test atmosphere, the atmosphere structure and abundances of the W7 deflagration model \citep{nomoto84}
are used.
The atmosphere structure is expanded to a point in time of 10 days after the explosion. The densities and radii are
determined by free homologous expansion and can be computed easily.
To perform the test calculations, we obtained an initial temperature structure with the \phx\ temperature
correction procedure \citep{phhtc03}. This temperature structure is the result of a simple approach
that does not represent the temperature structure of an SN Ia precisely. However, for the test calculations for 
our energy solver, this simple temperature structure is sufficient.
With this initial atmosphere structure, the energy solver is applied for different test cases.
All contributions to the energy change are tested separately.

\subsection{Energy transport}

In this section, the energy transport through the atmosphere is tested. The energy solver
considers only the energy change caused by emission and absorption of radiation, where
the result of the radiative transfer equation is needed. All other influences are neglected.
As a first test, we check how the initial temperature structure
changes if the energy solver 
is changing the SN Ia envelope.
As the initial atmosphere structure is already in radiative
equilibrium, the energy solver should not 
change the temperature structure significantly, because it also pushes the atmosphere towards a radiative equilibrium
state.

\begin{figure}
 \resizebox{\hsize}{!}{\includegraphics{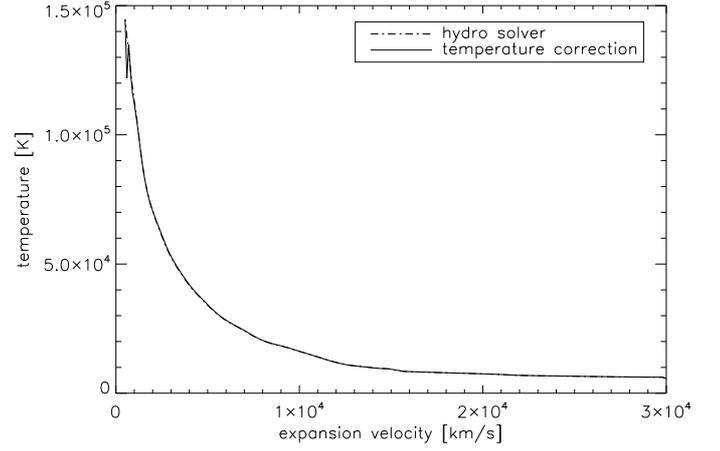}}
\caption{Temperature structures of a test atmosphere obtained with the energy solver and the \phx\ temperature correction procedure.}
\label{fig:lc_test_tcor}
\end{figure}

In Fig. \ref{fig:lc_test_tcor}, a comparison of the temperature structure of the energy solver to
the result of the temperature correction procedure is shown.
The differences in the temperature structure for most layers are less than 1\%.
But the temperature differences of the inner layers are clearly higher.
These differences arise in the \phx\ temperature correction result which produces a spike in the temperature structure. This
is likely due to the boundary condition in the temperature correction.
Hence, the resulting temperature structure obtained with the energy solver is more accurate.
Here, the temperature structure is smooth.
In order to obtain an atmosphere in radiative equilibrium, the energy transport part of the
energy solver can be used instead of the temperature correction procedure.
The main problem is that about a few hundred time steps are needed to obtain the resulting atmosphere structure
in radiative equilibrium,
while the temperature correction needs fewer iteration steps and is, therefore, significantly faster.

\subsection{Expansion}
In a further test calculation, the expansion part of the energy solver is checked.
The only energy change considered is the adiabatic cooling due to free expansion of the SN Ia envelope.
The energy deposition by $\gamma$-rays or an energy change due to energy transport is disabled.
For this test case, the expectation is that the atmosphere should just cool down, so the temperature of the atmosphere
and the observed luminosity should be decreasing.

\begin{figure}
 \resizebox{\hsize}{!}{\includegraphics{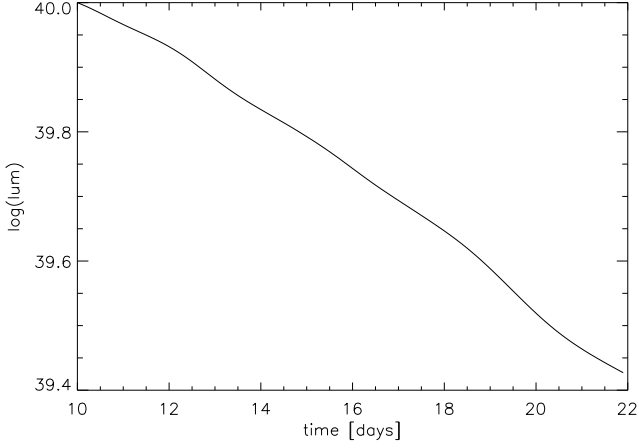}}
 \caption{Light curve of a test atmosphere that is just expanding and adiabatically cooling.}
 \label{fig:lc_expan_lum}
\end{figure}
\begin{figure}
 \resizebox{\hsize}{!}{\includegraphics{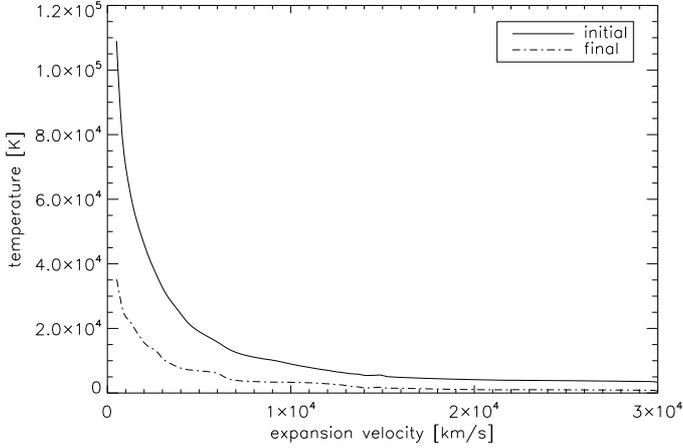}}
\caption{Initial and final temperature structure of a test atmosphere
which is just expanding and adiabatically cooling.}
 \label{fig:lc_expan_temp}
\end{figure}

The observed luminosity is shown in Fig. \ref{fig:lc_expan_lum}.
The observed luminosity of the SN Ia atmosphere decreases in time.
The temperature structure of the first and the last time step is plotted in
Fig. \ref{fig:lc_expan_temp}. The adiabatic expansion has cooled the atmosphere everywhere,
and the new temperature structure is now significantly lower than the initial one.

\subsection{Energy deposition}

To test the energy deposition by the radioactive decay of nickel and cobalt into the SN Ia atmosphere, a test case is
considered, where only this $\gamma$-ray deposition is calculated with the energy solver.
The energy change due to free expansion and energy transport are
neglected to see the direct effect of the additional energy put into the test SN Ia atmosphere.

\begin{figure}
 \resizebox{\hsize}{!}{\includegraphics[width=0.49\textwidth]{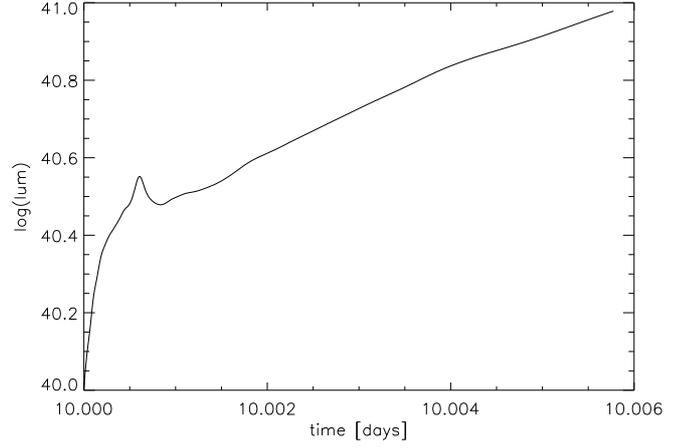}}
\caption{Light curve of a test atmosphere with energy deposition.
Because of the additional energy, the luminosity increases with time.}
 \label{fig:lc_gamma_lum}
\end{figure}

The results of the energy deposition by $\gamma$-rays  calculation with the energy solver
is shown in Fig. \ref{fig:lc_gamma_lum}, where the light curve is plotted.
Due to the energy added to the atmosphere, the luminosity seen by an observer increases with time.

\subsection{Realistic test scenario}

After all single effects have been tested, we now calculate a test case where all effects are considered for the solution of the energy solver.
Again, the same initial temperature structure is used and we start our calculation at day 10 after the explosion.
Free expansion as well as energy deposition and energy transport are active for this computation.

\begin{figure}
\resizebox{\hsize}{!}{\includegraphics[width=0.49\textwidth]{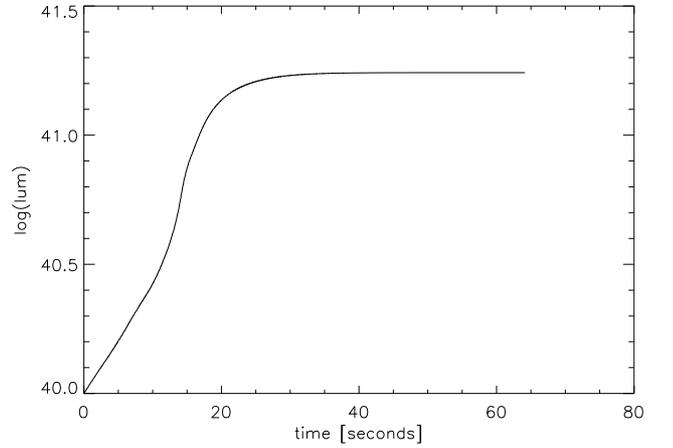}}
 \caption{Model light curve of a realistic test scenario.}
 \label{fig:lc_radeq_lum}
\end{figure}

\begin{figure}
 \resizebox{\hsize}{!}{\includegraphics[width=0.49\textwidth]{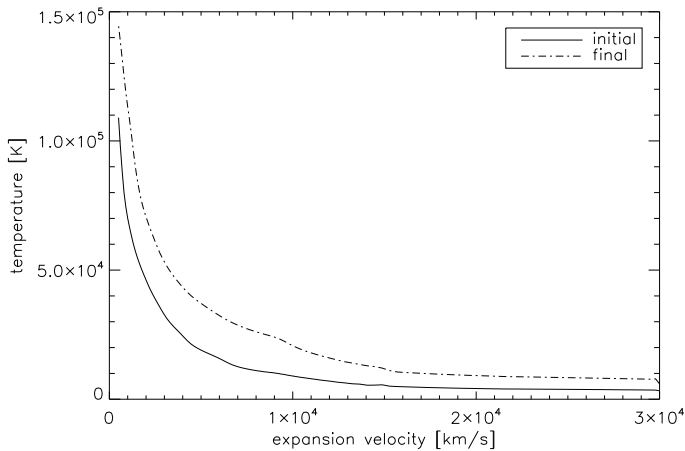}}
\caption{The initial temperature structure for the
    static solution of the equation of radiative equilibrium and the
    relaxed temperature structure obtained 1 minute later by advancing
    the energy solver in the 
  test scenario. 
All influences on the SN Ia atmosphere are considered.}
 \label{fig:lc_radeq_temp}
\end{figure}

The observed light curve is shown in Fig. \ref{fig:lc_radeq_lum}.
The luminosity increases because of the energy input from deposition
of energy from the 
$\gamma$-rays produced by  decay of
nickel and cobalt.
It takes a certain time, until the whole atmosphere has relaxed to this new condition. The atmosphere is then in
radiative equilibrium state, and the luminosity stays constant. The initial and final temperature structure are plotted
in Fig. \ref{fig:lc_radeq_temp}. The energy input caused by radioactive decay has increased
the temperature of the whole atmosphere.
The atmosphere is heated by $\gamma$-ray deposition in the inner part of the atmosphere.
Due to this additional energy, the luminosity of these
layers increases and the heat is radiated away and absorbed by the surrounding layers. This energy transport
takes care that the deposited energy is moving through the whole atmosphere so that the
temperature increases everywhere and
the additional energy from the radioactive decay is radiated away towards the observer.
The atmosphere is then in radiative equilibrium.

\section{SN Ia model light curves}

We now present our first theoretical SN Ia light curves obtained with our extensions
to the general purpose model atmosphere code \phx\ and compare them to observed SN Ia light curves.
The online supernova spectrum archive (SUSPECT) \citep{suspect01,suspect02}\footnote{http://suspect.nhn.ou.edu/} provides numerous of observations
of different types of supernovae.
For this work, the observed light curves of SN 2002bo and SN 1999ee are used to compare them to our
results of model light curves.
Photometric light curve observations of SN 2002bo in different photometric bands \citep{benetti04} have been obtained.
SN 1999ee also has observed spectra \citep{hamuy99ee99ex02} and photometry \citep{stritz99ee99ex02}.

\subsection{Method}

\label{sec:method}
The energy solver is now applied to calculate synthetic light curves of SNe Ia.
The SN Ia light curve evolution is calculated during the free expansion phase.
For the initial model atmosphere structure, the results of the
explosion calculation of other groups are used as the input structure. Each layer has a certain
expansion velocity, which does not change during the evolution, because homologous expansion is assumed.
We start the model light curve calculation a few days after explosion. In the first few days the
SN Ia envelope is optically thick and compact.
Another point is, in the first few days the SNe Ia light curves are quite faint, and there are almost
no observations of this early time phase that have been obtained. Thus, for the model light curve calculation,
it is adequate to start the light curve calculation a few days after the explosion.
The initial structure is given by the result of the explosion model simulation.
The results of the explosion model give the expansion velocities, density structure and the
non-homogeneous abundances of all chemical elements present in the SN Ia envelope.
The envelope expands for the first few days by assuming homologous expansion.
The radii are determined by the expansion velocities and time after explosion.
For the first temperature structure guess, the \phx\ temperature correction procedure is used to obtain a initial
temperature structure, which is in radiative equilibrium. Alternatively we could calculate RE models starting
at day one.

For the computation of an SN Ia model light curve, we let the energy solver change
the obtained initial model atmosphere structure.
The atmosphere structure adapts to the new conditions caused by
$\gamma$-ray deposition and other energy effects. After a certain time, the atmosphere eventually reaches
the radiative equilibrium state.
A typical time step of a energy change in the model atmosphere is about $0.1$s.
It takes about 500 time steps to reach radiative equilibrium depending on in which evolution phase the SN Ia is.
For the later phase after the maximum of the SN Ia light curve, fewer time steps are needed.
We use this first radiative equilibrium atmosphere structure to calculate a more detailed spectrum,
where more wavelength points are used.
This model spectrum is then used to obtain the first point of the model light curve for each band
by using the filter functions described below.
It certainly would require too much computation time
to calculate a whole light curve evolution by using the typical time steps of about $0.1$s.
Therefore, to obtain the next point of the light curve, big time steps are computed. For these big time steps,
the atmosphere is only expanding.
This means that neither energy deposition by $\gamma$-rays  or energy transport through the atmosphere
is considered for the solution of the energy solver.
After half a day computed with big time steps,
the whole energy solver changes the atmosphere structure again and the next point in the light curve
is obtained after the atmosphere structure moves back to radiative equilibrium.
Even with the use of these big time steps, the radiative transfer for
a light curve of 50 days has to be solved about 10,000 times. For normal model atmosphere calculation in \phx\
with the temperature correction procedure, the radiative transfer equation has to be solved about 100 times, at most.
Therefore, the computation of a whole SN Ia model light curve is very computationally expensive,
because the solution of the radiative transfer equation has to be obtained too many times. It takes
a huge amount of computation time to calculate even simple SN Ia model light curves,
even when using only a few wavelength points.
Basically, the SN Ia model light curve is a curve consisting of half day points, where the atmosphere is in radiative equilibrium.
During the later phase after maximum light, the big time steps have been performed for one or even two days, as the energy changes is the SN Ia atmosphere become smaller.

\begin{figure}
\centering
\resizebox{\hsize}{!}{\includegraphics{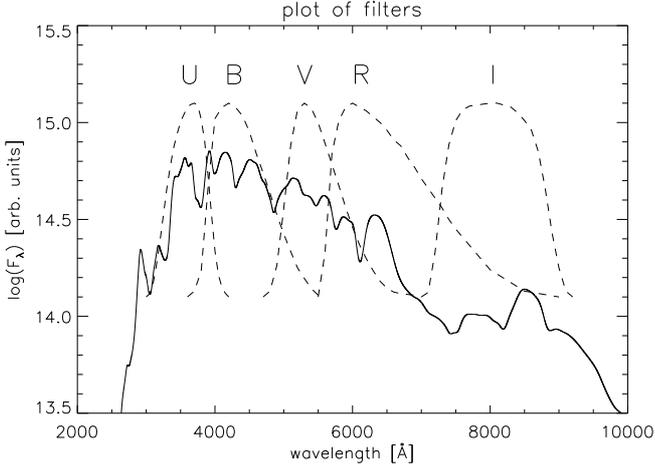}}
  \caption{A typical model spectrum of an SN Ia together with the filter functions for the U, B, V, R and I band.
These are used to determine the luminosity in each band for the model light curves.}
  \label{fig:filters}
\end{figure}

As described above, a model spectrum is calculated with the obtained atmosphere structure, which is
in radiative equilibrium at each half day point of the light curve.
To obtain a model light curve for different photometric bands,
filter functions were used to calculate the luminosity in different bands.
In Fig. \ref{fig:filters}, an SN Ia model spectrum and the filter
functions for the  U, B, V, R and I bands are shown. These filter functions are described in \citet{hamuyetal92}.
With these filter functions, SNe Ia model light curves can now be obtained for these five different photometric bands.

\subsection{Light curves of LTE models}

For the first calculations of theoretical light curves, the model atmosphere of the SN Ia is considered to be in LTE.
For a first approach to obtain model light curves
this is adequate.
Another reason for the assumption of LTE is that model light curve calculations with an atmosphere treated in NLTE
use more computation time.
In order to obtain an SN Ia model light curve in reasonable computation time, a model atmosphere in LTE is a necessary assumption.
For the model light curve calculations presented in this section the following parameters were chosen.
The model atmosphere is divided into 128 layers.
The radiative transfer is solved including atomic lines of the Kurucz atomic data line list.
The number of wavelength points used for the solution of the radiative transfer is about 2400.
For the inner boundary condition of the radiation, we use the nebular boundary condition,
which means that the radiation is just passing through the inner empty region,
therefore all the radiation is produced by the ejecta itself, there is no inner lightbulb.
The outer boundary condition is zero.

For the first LTE light curve calculation, the W7 deflagration explosion model is used
to obtain model light curves of SNe Ia in different photometric bands.
The first point of the light curve was calculated at three days after the explosion.
The method to obtain the theoretical light curves is described in section \ref{sec:method}.

\begin{figure}
 \resizebox{\hsize}{!}{\includegraphics{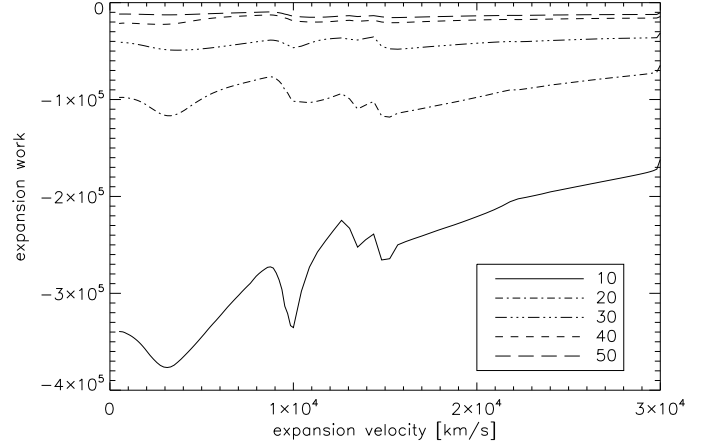}}
\caption{The energy change due to the expansion for several days.}
 \label{fig:lc_expan_dt}
\end{figure}
\begin{figure}
 \resizebox{\hsize}{!}{\includegraphics{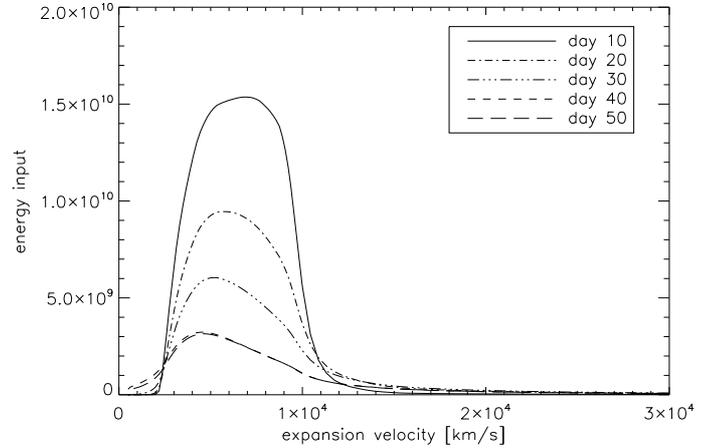}}
\caption{The energy change due to the additional energy input for several days.}
 \label{fig:lc_source_dt}
\end{figure}
\begin{figure}
 \resizebox{\hsize}{!}{\includegraphics{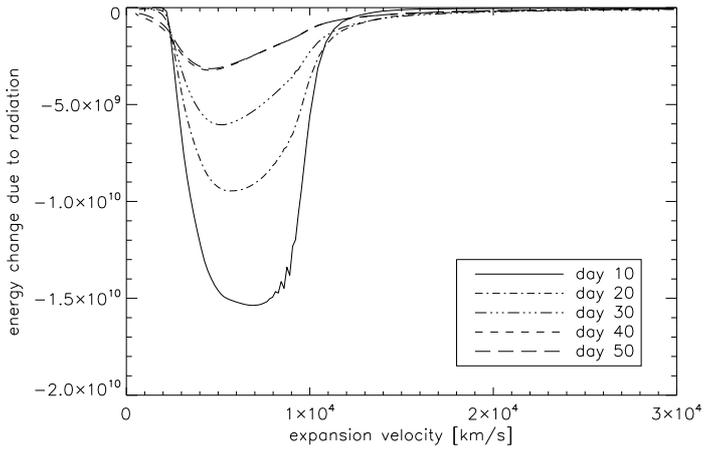}}
\caption{The energy change due to the radiation for several days.}
 \label{fig:lc_rad_dt}
\end{figure}
\begin{figure}
 \resizebox{\hsize}{!}{\includegraphics{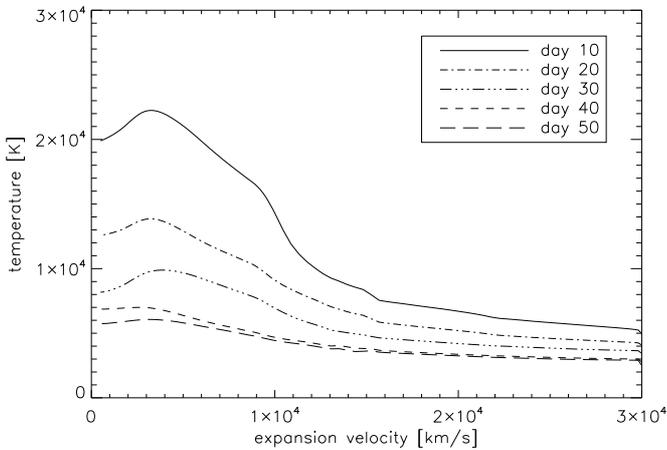}}
\caption{Temperature profiles as a function of time.}
 \label{fig:lc_temp_dt}
\end{figure}

Before we present our resulting model light curves, we present plots of various terms that change the 
energy density for several moments in time. In Fig. \ref{fig:lc_expan_dt}, the energy change due to the expansion 
is shown for several moments in time. The energy change is decreasing for later days in the light evolution.
The additional energy input due to the radioactive decay is shown in Fig. \ref{fig:lc_source_dt}.
With increasing time, the amount of energy input into the atmosphere decreases. It also shows, that the additional energy 
is located in the inner part of the envelope, where the $^{56}$Ni has been produced.
Fig. \ref{fig:lc_rad_dt} shows the energy loss due to the emission of radiation. 
The temperature structures of the atmosphere for several moments in time are shown in Fig. \ref{fig:lc_temp_dt}.

\begin{figure}
 \centering
 \resizebox{\hsize}{!}{\includegraphics{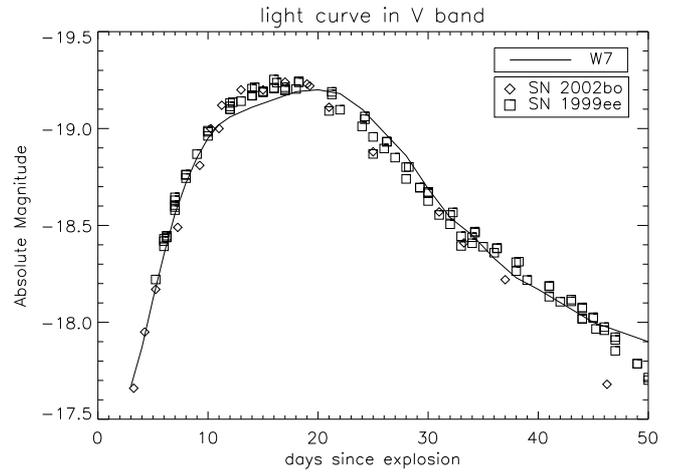}}
 \caption{LTE model light curve of the W7 explosion model in the V band compared to two observed SN Ia
light curves of SN 1999ee and SN 2002bo.}
 \label{fig:lc_v_neb}
\end{figure}

Figure \ref{fig:lc_v_neb} shows the LTE SN Ia model light curve of the W7-based explosion model in the optical V band.
The theoretical light curve accurately reproduces the observed light curves of two SNe Ia.
The steep rise of the model light curve beginning at three days after explosion is in agreement with the observed
light curves. The maximum of the W7-based model light curve seems to be later than that of the observed light curves.
At 20 days after the explosion, the model light curve has its maximum, while the maximum of the observed light curves
is around 17 days after the explosion. After maximum, the decline of the light curve of the W7-based
model well reproduces the observed light curve.
Even up to the later phase at 50 days after the explosion, where the atmosphere becomes significantly thinner,
the fit to the observed light curves is quite accurate.

\begin{figure}
 \resizebox{\hsize}{!}{\includegraphics{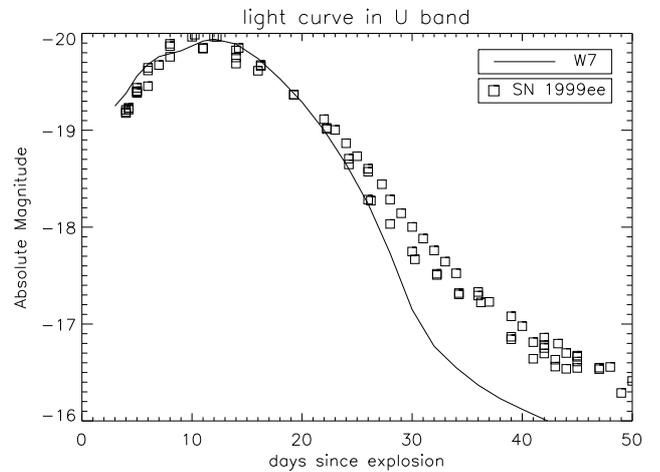}}
 \caption{U band light curve of the LTE model light curve compared to SN 1999ee.}
 \label{fig:lc_u}
\end{figure}
\begin{figure}
 \resizebox{\hsize}{!}{\includegraphics{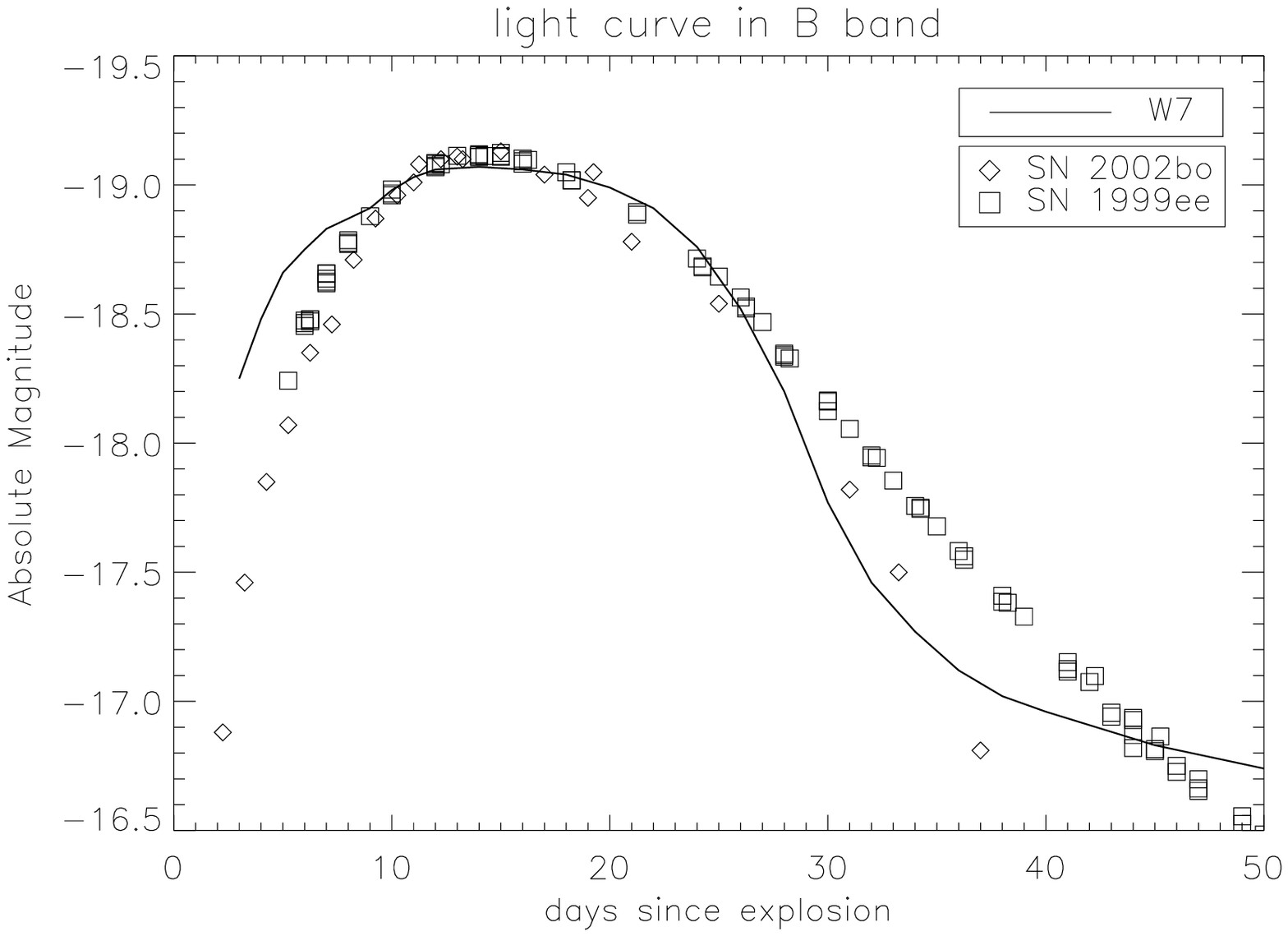}}
 \caption{LTE light curves of the W7 explosion model. The B band model light curve
is too bright during the first few days.}
 \label{fig:lc_b}
\end{figure}

The theoretical light curve in the ultraviolet U band is shown in Fig. \ref{fig:lc_u}.
Only an observed light curve of SN 1999ee is available.
The observational data are scattered.
The rise in the beginning as well as the maximum phase is well represented by the model light curve.
However, the decline of the theoretical light curve seems to be too steep. This same effect is present
in the model light curve of the B band, which is shown in Fig. \ref{fig:lc_b}. The
first days of the model light curve are too bright compared to both observed SN Ia light curves. The
maximum phase of the model light curve is in good agreement with the observed ones. At day 50, the model light
curve becomes brighter than the observed light curves.

\begin{figure}
 \resizebox{\hsize}{!}{\includegraphics{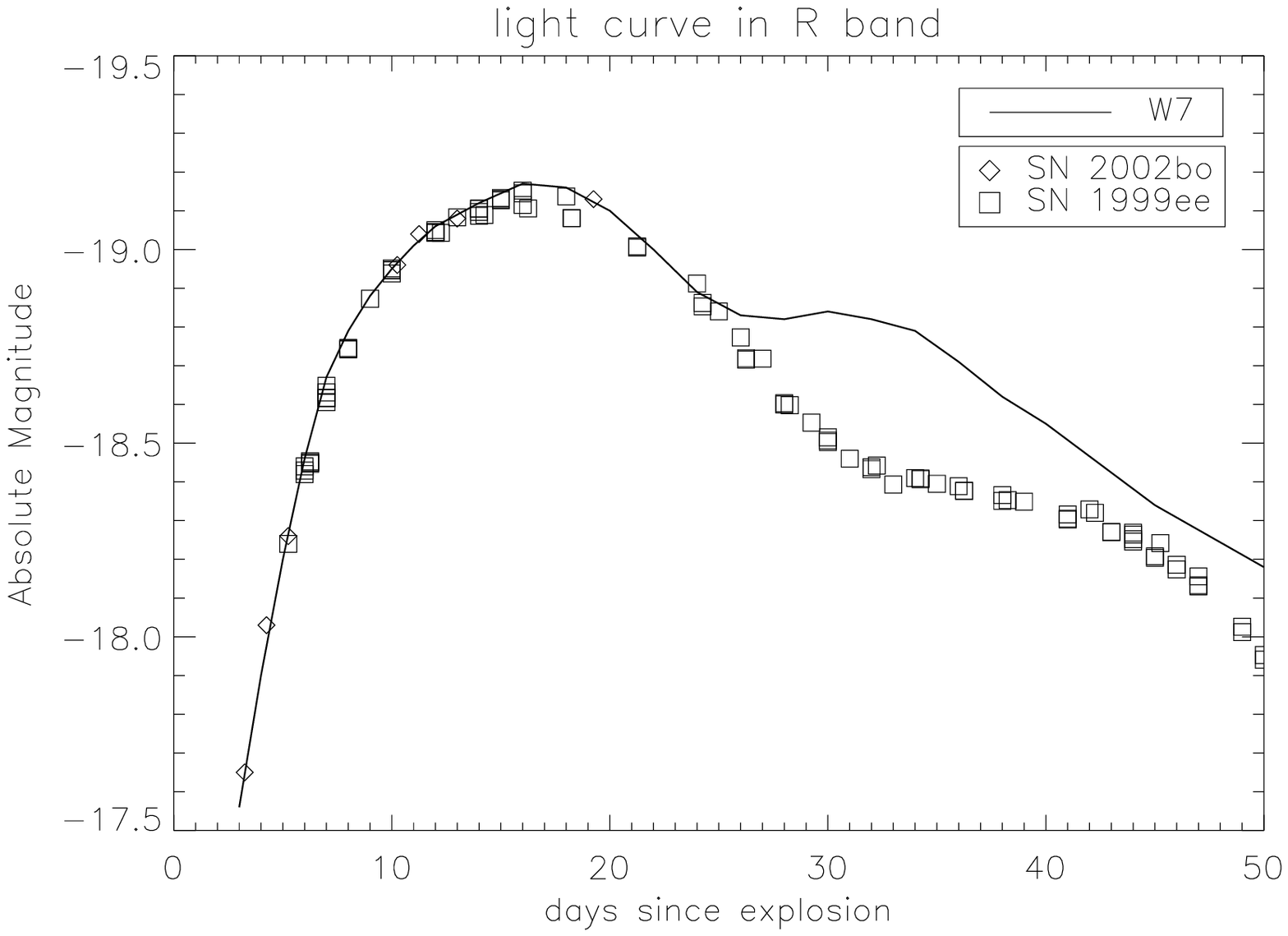}}
 \caption{The theoretical light curve in the R band
seems to rise after a first decline after the maximum phase.}
 \label{fig:lc_r}
\end{figure}

\begin{figure}
 \resizebox{\hsize}{!}{\includegraphics{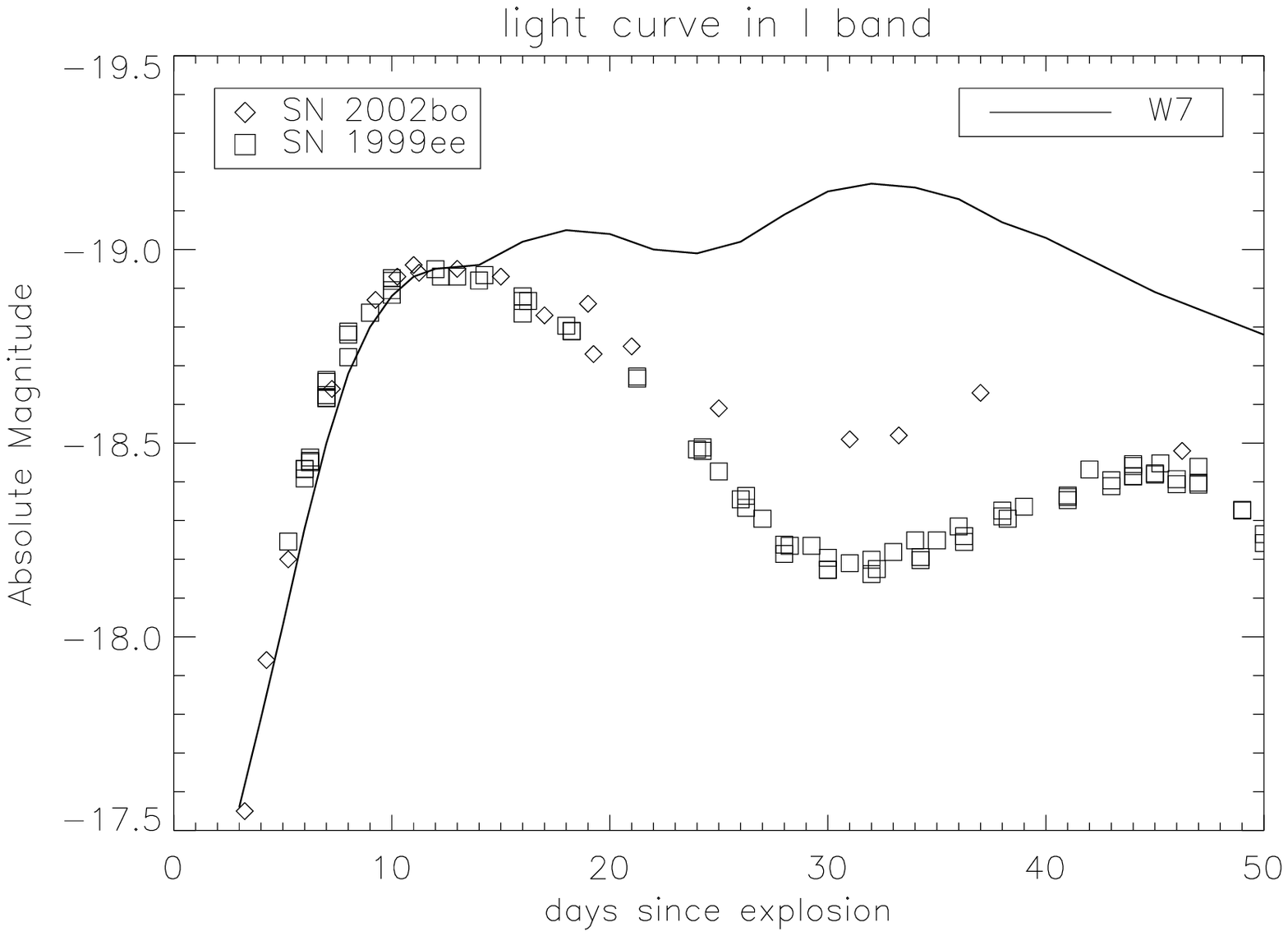}}
 \caption{LTE light curves of the W7 explosion model.
For the I band the model light curve has no distinctive maximum.
At 30 days after the explosion the model light curve is way too bright in comparison to the observed ones.}
 \label{fig:lc_i}
\end{figure}

In Fig. \ref{fig:lc_r}, a plot of the model light curve of the R band is shown.
The steep rise in the beginning and the maximum phase of the observed SN Ia light curves is
well represented by the computed model light curve.
However,
the theoretical light curve fit becomes worse for later phases.
The luminosity of the model light curve seems to
rise again at around day 25 after the explosion.
Up to day 45, the model light curve has a second bump, which is not observed in the light curves of SN 1999ee and SN 2002bo.
In the infrared I band, the decline
after the maximum phase is missing, as shown in Fig. \ref{fig:lc_i}.
As in the R band, the rise in the beginning and maximum are well represented in the model light curve.
However, at maximum, the luminosity of the SN Ia model light curve rises further,
which is not seen in the observed light curves of SN 2002bo and 1999ee.
Around day 30, the difference between model and observed light curve in the I band are about 1 mag.
Up to day 50, the model light curve declines, while the observed light curves show their second maximum around 40 days
after explosion.

\begin{figure}
 \resizebox{\hsize}{!}{\includegraphics{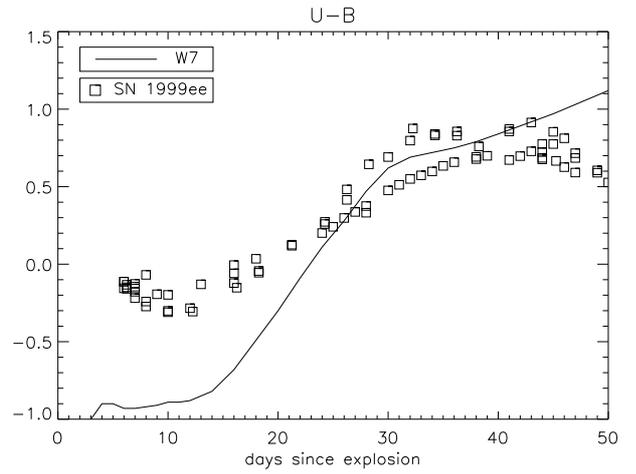}}
 \caption{Comparison of U-B for the theoretical and observed light curves.}
 \label{fig:lc_u-b}
\end{figure}
\begin{figure}
 \resizebox{\hsize}{!}{\includegraphics{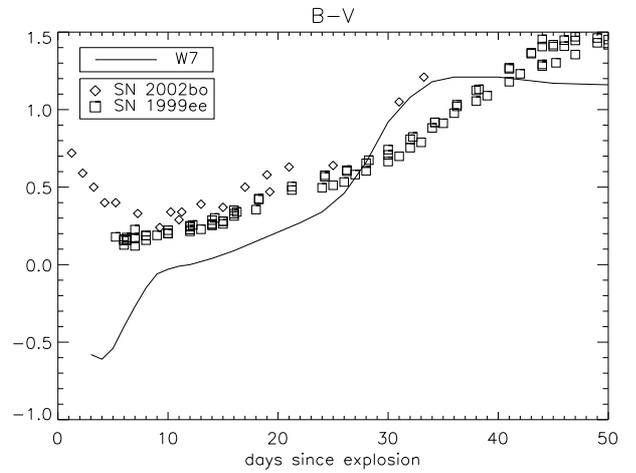}}
 \caption{Comparison of B-V for the theoretical and observed light curves.}
 \label{fig:lc_b-v}
\end{figure}

In Fig. \ref{fig:lc_u-b}, we compared the relation U-B of the theoretical and observed light curves.
The same comparision for the B-V relation is shown in Fig. \ref{fig:lc_b-v}.

\subsection{Dynamical models}

We have three different results of explosion calculations of SN Ia events.
The structure of these models are results from hydrodynamical
explosion calculations.
One is the W7 deflagration model of \citet{nomw7}. The W7 dynamical model assumes that the ongoing
explosion is a deflagration. The flame propagates with a velocity lower than the speed of sound
outwards.
Another possible explosion model is the delayed detonation model. The explosion starts with a
deflagration which eventually proceeds to a detonation. We used two different dynamical models
named as DD25 and DD16. For more detailed information about the used delayed detonation
models see \citet{HGFS99by02}.

\begin{figure}
\resizebox{\hsize}{!}{\includegraphics{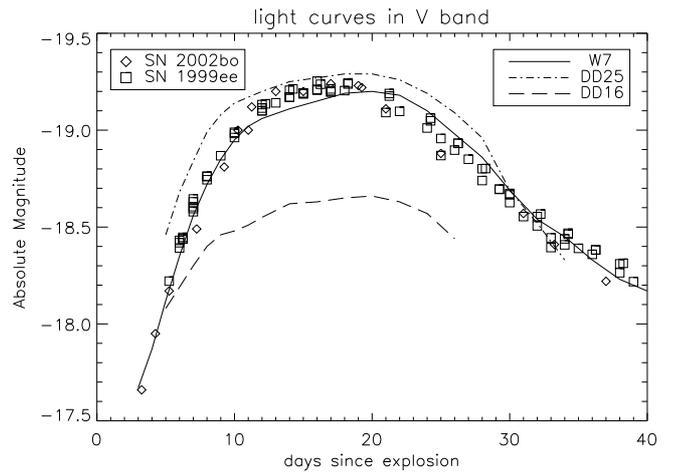}}
  \caption{V band light curves for three different dynamical models W7, DD16 and DD25.
 The DD25 dynamical model is fainter because less nickel was produced during
the explosion. The W7 and DD15 dynamical model light curves are quite similar.}
  \label{fig:lc_hyd}
\end{figure}

The resulting optical model light curves of the V band are plotted in Fig. \ref{fig:lc_hyd}.
The light curve of the DD16 is fainter than the other ones.
This is due to the fact that in this explosion model less $^{56}$Ni is produced.
The W7 and DD25 model are quite similar. DD25 is slightly brighter than W7.
The shape of the light curves are in good agreement with the observed ones.
The maximum of the DD16 is later than the one of the W7 and DD25.

\begin{figure}
\resizebox{\hsize}{!}{\includegraphics{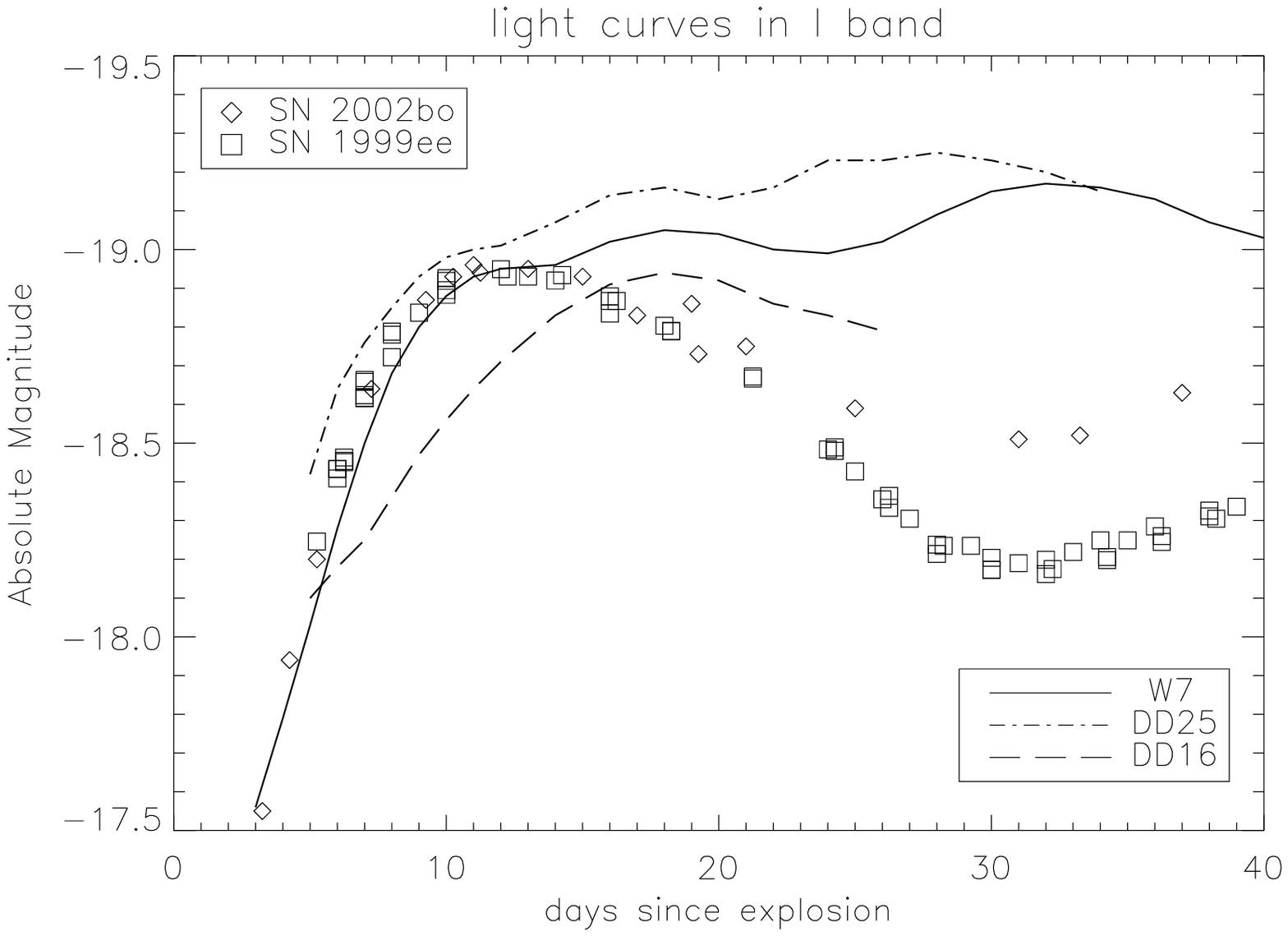}}
  \caption{I band light curves for the dynamical models W7, DD16 and DD25 are compared to the two observed
SN Ia light curves.}
  \label{fig:lc_hyd_i}
\end{figure}

In Fig. \ref{fig:lc_hyd_i} the light curves in the I band of the three dynamical models are shown.
The infrared light curves deviate strongly from the observations.
As in the optical light curve, the
DD16 light curve is fainter than W7 and DD25.
DD16 also has only one maximum and a decline. However, this maximum is
later than the observed ones. The DD25 and W7 light curves are still rising
after the observed light curves have reached their maximum and are declining.
The theoretical light curves in the infrared must be improved for all explosion models.

\section{Light curves of NLTE models}

The LTE model light curves in the V Band and most other bands are in quite good agreement with observed SN Ia light curves.
However, in the near-infrared of the I band, the model light curves need improvements especially for the
later phase to fit the observational light curves more accurately.
So far, the SN Ia model atmosphere is considered to be in LTE during the whole evolution time.
In this section, we calculate the model light curves with the assumption of an atmosphere
which is not in LTE. The computation of model light curves with SN Ia model atmospheres in NLTE
requires a huge amount of computation time.
At first, we model light curves with atmospheres in NLTE that are computed with LTE temperature structures.
More realistic NLTE model light curves with a temperature structure that
adapts to NLTE conditions are computed to investigate the NLTE effects on the model light curves.

The first approach to compute NLTE model light curves is to consider the atmosphere to be in NLTE, but use a fixed LTE temperature structure.
For this computation of NLTE model light curves, we used the calculated radiative equilibrium LTE temperature structure
of the W7 deflagration model.
We keep this temperature structure constant and perform 20 iterations to let the NLTE converge, which is mainly
the occupation numbers of the species that are considered for the NLTE calculation.
We considered the following species for the calculations in NLTE:
H~I, He~I, He~II, C~I-III, O~I-III, Ne~I, Na~I, Mg~I-III, Si~I-III, S~I-III, Ca~II, Ti~II, Fe~I-III and  Co~II.
These are the species that are most abundant in SN Ia atmospheres and mainly contribute to the spectrum.
There is only little H and He in SN Ia, and in W7, there is none. However, considering H~I, He~I and He~II in NLTE does not cost much
computation time.
The advantage of this approach to an NLTE model light curve is that no temperature iterations have to be performed.
This reduces the computation time significantly, although about 200,000 wavelength points are calculated
instead of 2,400 in case of LTE.

\begin{figure}
 \centering
 \resizebox{\hsize}{!}{\includegraphics{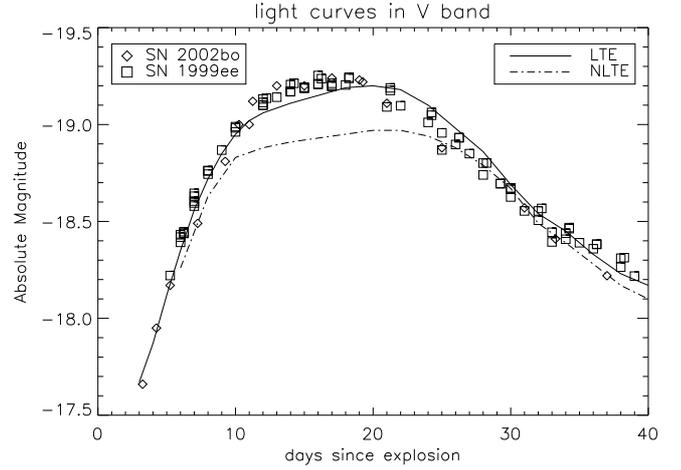}}
 \caption{Model light curves of the W7 explosion model in the V band. The NLTE model atmosphere has
an LTE temperature structure.}
 \label{fig:lc_nltefixv}
\end{figure}
In Fig. \ref{fig:lc_nltefixv}, the SN Ia model light curve in the V band of an NLTE calculation with an LTE temperature
structure is shown.
The NLTE light curve is fainter than the LTE light curve. During the maximum phase there is about 0.4 mag
difference between both light curves. We have to point out that for our NLTE models, the energy is not conserved
because the radiation does not thermalize within the envelope.
After maximum the NLTE model light curve approaches the LTE light curve
and agrees with the observed light curves as well as the LTE light curve.
However, the NLTE model light curve in the V band does not agree with the observed light curves very accurately.
With the assumption of an atmosphere in LTE, a better fit is obtained, although NLTE is more accurate.

\begin{figure}
 \resizebox{\hsize}{!}{\includegraphics{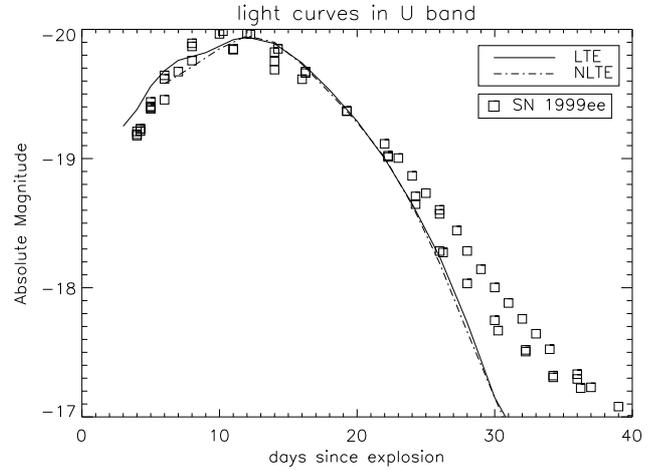}}
 \caption{Model light curves of the W7 model in the U band. The NLTE model atmosphere has an LTE temperature structure.}
\label{fig:lc_nltefixu}
\end{figure}

\begin{figure}
 \resizebox{\hsize}{!}{\includegraphics{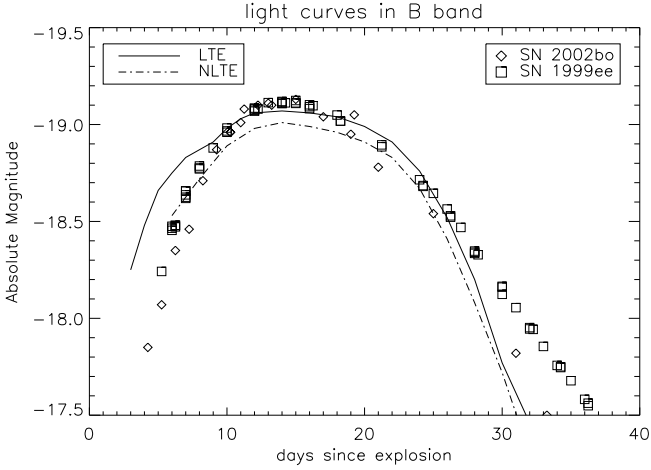}}
 \caption{Model light curves of the W7 model in the B band. The NLTE model atmosphere has an LTE temperature structure.}
 \label{fig:lc_nltefixb}
\end{figure}
The NLTE model light curve of the U band is shown in Fig. \ref{fig:lc_nltefixu}. The NLTE light curve shows almost no
deviations from the LTE light curve. 
During the later
phase the steep decline is also present in the NLTE model light curves.
In Fig. \ref{fig:lc_nltefixb}, the NLTE model light curve in the B band is presented. 
The NLTE model light curve is slightly fainter than the LTE light curve. 
The shape of the NLTE light curve seems to be the same as for the LTE light curve.
The NLTE model light curve is also an accurate fit to the observed
light curves.

\begin{figure}
 \resizebox{\hsize}{!}{\includegraphics{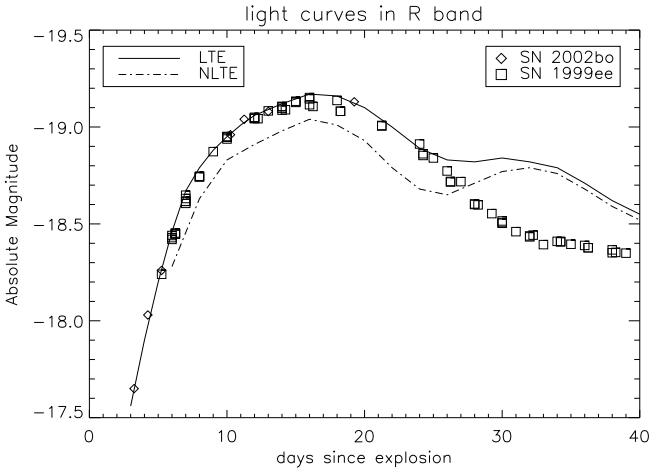}}
 \caption{NLTE and LTE model light curves in the R band of the W7 model.
The NLTE model atmosphere has an LTE temperature structure.}
 \label{fig:lc_nltefixr}
\end{figure}
\begin{figure}
 \resizebox{\hsize}{!}{\includegraphics{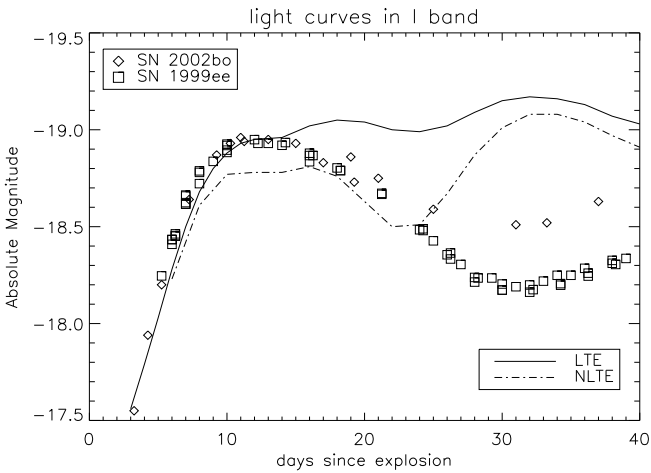}}
 \caption{NLTE and LTE model light curves of the W7 model. The NLTE model atmosphere has an LTE temperature structure.
In the I band the NLTE light curve is somewhat of an improvement.}
 \label{fig:lc_nltefixi}
\end{figure}
In Fig. \ref{fig:lc_nltefixr}, the NLTE model light curve of the R band is shown. The luminosity of the NLTE light curve is fainter
than the LTE light curve. It has almost the same shape. But for the phase after 25 days after the explosion,
the NLTE model light curve rises again.
The assumption of NLTE does not improve the fit to the observed light curves of SN 1999ee and SN 2002bo.
The infrared NLTE light curve for the I band is shown in Fig. \ref{fig:lc_nltefixi}. During the maximum phase, the NLTE light
curve is fainter than the LTE light curve. A distinctive maximum is missing in the NLTE model light curve.
For the phase between day 15 and day 25, the NLTE model light curve fits the observed SN Ia
light curves very accurately. Here, there is a significant improvement compared to the light curve of the LTE model atmosphere.
But, after day 25 the NLTE model light curve starts to rise and becomes too bright.
This is the same problem that already emerged in the LTE light curve.

\subsection{NLTE atmosphere structures}

We perform a more realistic NLTE calculation of the SN Ia model atmosphere evolution.
The temperature structure is now changing and adapting to the new conditions of NLTE.
The calculation of an NLTE model light curve takes a huge amount of computation time.
Considerably more wavelength points are needed for the calculation of
the solution of the radiative transfer.
For all the species considered in NLTE, the rate equations have to be solved.
Note that a time step in the NLTE calculation is not a real time step.
The rate equation changes the energy of the atmosphere, but this
is not included in the energy solver. However, it is adequate as the goal is to obtain a temperature structure,
where the atmosphere is in radiative equilibrium.

In a first try to calculate a more realistic NLTE light curve, numerous species up to calcium are considered to be in NLTE.
These are the species H~I, He~I, He~II, C~I-III, O~I-III, Ne~I, Na~I, Mg~I-III, Si~I-III, S~I-III and Ca~II.
Higher species are neglected because they have more
levels, which would increase the computation time significantly.
Nevertheless, the computation of an NLTE light curve needs significantly more time than LTE. For this computation
of NLTE model light curves, about 200,000 wavelength points need to be calculated, compared to about 2,400 wavelength points for a pure LTE light curve calculation.
To obtain an NLTE model light curve in a reasonable time, we started the calculation at day 10 after the explosion
and used the LTE temperature structure as the initial atmosphere structure.
The main focus is to check if the infrared light curves during the later phase can be further improved.

\begin{figure}
 \resizebox{\hsize}{!}{\includegraphics{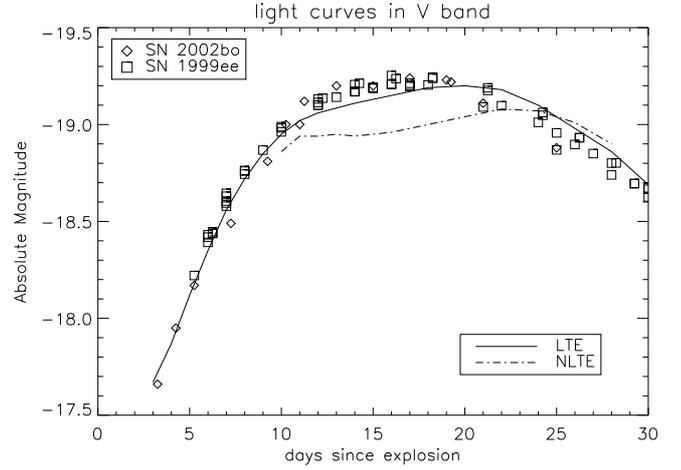}}
 \caption{NLTE and LTE model light curves of the W7 explosion model in the V band.}
 \label{fig:lc_nltev}
\end{figure}
\begin{figure}
 \resizebox{\hsize}{!}{\includegraphics{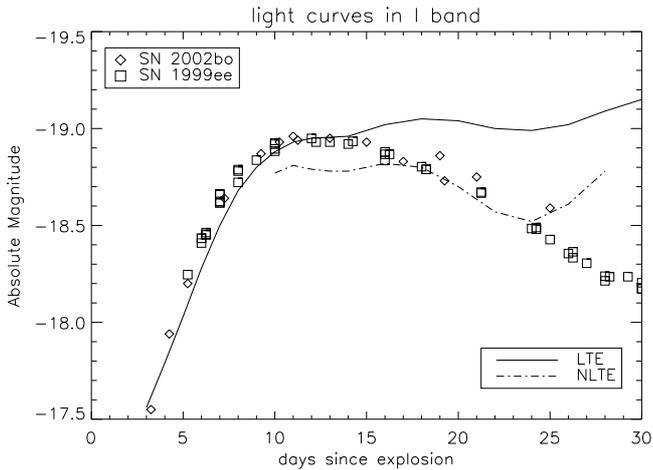}}
 \caption{NLTE and LTE model light curves of the W7 explosion model in the I band.}
 \label{fig:lc_nltei}
\end{figure}
In Fig. \ref{fig:lc_nltev}, the NLTE and LTE light curves of the W7 deflagration model in the V band are shown.
The maximum phase is not well fitted by the NLTE model light curve. The LTE light curve fits the
observed light curves better. At day 20 the NLTE and LTE light curves are almost the same.
Compared to the NLTE light curve obtained with the LTE temperature structure, there are only small differences to the
NLTE calculation where the temperature structure adapts to the NLTE conditions.

In Fig. \ref{fig:lc_nltei}, the I band model light curves of NLTE and LTE are shown.
During the maximum phase, the NLTE light curve is fainter than the LTE light curve.
Between day 15 and day 25 the NLTE model light curve fits the observed light curves quite accurately.
Here, the use of NLTE improves the fit to the observed light curves. However, we obtained this improvement also
with the NLTE model light curve calculated with the LTE temperature structure.
At day 25 the NLTE model light curve starts to rise again.
Although the consideration of NLTE improves the model light curve in the I band,
the problem with a rise in brightness after the maximum remains.

\subsection{Line scattering}

\begin{figure}
 \resizebox{\hsize}{!}{\includegraphics{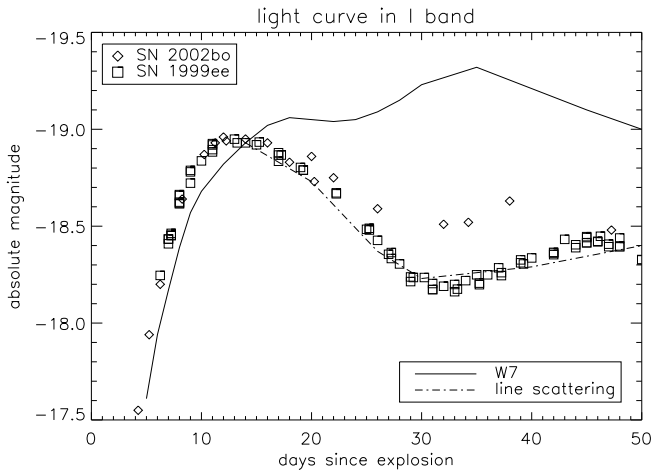}}
 \caption{Light curves in the I band. The light curve with variable scattering is fainter than the LTE light curve.}
 \label{fig:lc_iband}
\end{figure}

The fits to the infrared light curves need to be improved, as they are
too bright during the phase of 20 to 50 days after the
explosion in all models. We
found that a change in the ratio of line scattering to absorption in the LTE
case improves the fits to the I band light curve.  The resulting light curve is
shown in Fig. \ref{fig:lc_iband}.  The line scattering is here roughly
inversely proportional to the density and was for the figure ``hand tuned'' to
produce a good match.  As the density becomes lower due to the ongoing
expansion, the scattering becomes more important, as expected.  A more detailed
investigation of the IR line scattering and its effects on SN Ia light curve
modeling will be a topic of future work. To treat this effect more accurately,
very detailed NLTE model atoms are required that include the important
infrared lines.
In the atomic data we have presently available, these lines are mostly ``predicted''
lines and were not included in the our current  NLTE model atoms. This
result indicates that an accurate treatment of line scattering and NLTE 
effects in the IR lines is very important to model and interpret the IR light
curves of SNe.

\section{Conclusion}

We have presented a new approach for a energy solver which can be applied to expanding SN Ia envelopes
to compute SN Ia model light curves.
Test calculations confirm that the implemented code works properly.
We applied the energy solver to calculate SN Ia model light curves during the free expansion phase.

We present the first \phx\ light curves of type Ia supernovae in different photometric bands.
At first, we solved the radiative 
transfer equation with the assumption of the atmosphere being in LTE.
The model light curves were in decent agreement with the observed ones at early times.
However, in the infrared I band, the theoretical light curves do not fit the observed light curves.
\citet{blinn06a} had the same deviations in their model light curves compared to observations.
Their light curve in the I band is also too bright and further rising, even after the maximum in observed light curves.
\citet{kasen06b} presented a detailed study of SN Ia light curves in the near IR.
His I band light curve fits the observations better.
He also showed how variable and dependent on
different parameters the light curves in the near infrared can be.

Three different dynamical models were compared to observed
SN Ia light curves.
The delayed detonation model DD 16 is unlikely to be the best explosion model because the model light curves
are too faint to fit the observed light curves due to the low $^{56}$Ni mass.
The results of the W7 model are the best fit to the observed light curves.

For the later phase of the light curves, especially in the infrared, we need the
atmosphere to be considered in NLTE. We showed that NLTE model light curves computed
with an LTE temperature structure leads to improvements in the model light curves.
For future work more investigation of the infrared model light curves and the NLTE needs
to be performed. The focus has to lie on the influence of line scattering on the shape
of the light curves.
It would also be interesting to look in detail on the spectral evolution
during the free expansion phase.
Further improvements may be achieved with a full 3D radiative transfer and energy solver.

\begin{acknowledgements}
This work was supported in part by the
Deutsche Forschungsgemeinschaft (DFG) via the SFB 676, NSF grant AST-0707704,
and US DOE Grant DE-FG02-07ER41517.
This research used resources of the National Energy Research
Scientific Computing Center (NERSC), which is supported by the Office
of Science of the U.S.  Department of Energy under Contract
No. DE-AC02-05CH11231, and the H\"ochstleistungs Rechenzentrum Nord
(HLRN). We thank all these institutions for generous allocations of
computation time.
\end{acknowledgements}

\bibliographystyle{aa}
\bibliography{14778bib}

\end{document}